\documentclass[
 reprint,
 amsmath,amssymb,
 superscriptaddress,
 aps,
 prx,
 nofootinbib,
]{revtex4-2}

\usepackage{graphicx}
\usepackage{dcolumn}
\usepackage{bm}
\usepackage{physics}
\usepackage{siunitx}
\usepackage{appendix}
\usepackage[colorlinks,linkcolor=blue,urlcolor=blue, citecolor=blue]{hyperref}
\usepackage[whole]{bxcjkjatype}
\usepackage[T1]{fontenc} 
\usepackage{comment}

\begin{document}

\title{Scalable Networking of Neutral-Atom Qubits:\\
       Nanofiber-Based Approach for Multiprocessor Fault-Tolerant Quantum Computer}
\author{Shinichi Sunami}
\email{shinichi.sunami@nano-qt.com}
\affiliation{Nanofiber Quantum Technologies, Inc. (NanoQT), 1-22-3 Nishiwaseda, Shinjuku-ku, Tokyo 169-0051, Japan.}
\affiliation{Clarendon Laboratory, University of Oxford, Oxford OX1 3PU, United Kingdom}
\author{Shiro Tamiya}
\affiliation{Nanofiber Quantum Technologies, Inc. (NanoQT), 1-22-3 Nishiwaseda, Shinjuku-ku, Tokyo 169-0051, Japan.}
\author{Ryotaro Inoue}
\affiliation{Nanofiber Quantum Technologies, Inc. (NanoQT), 1-22-3 Nishiwaseda, Shinjuku-ku, Tokyo 169-0051, Japan.}
\author{Hayata Yamasaki}
\email{hayata.yamasaki@nano-qt.com}
\affiliation{Nanofiber Quantum Technologies, Inc. (NanoQT), 1-22-3 Nishiwaseda, Shinjuku-ku, Tokyo 169-0051, Japan.}
\affiliation{Department of Physics, Graduate School of Science, The Univerisity of Tokyo, 7-3-1 Hongo, Bunkyo-ku, Tokyo, 113-0033, Japan}
\author{Akihisa Goban}
\email{akihisa.goban@nano-qt.com}
\affiliation{Nanofiber Quantum Technologies, Inc. (NanoQT), 1-22-3 Nishiwaseda, Shinjuku-ku, Tokyo 169-0051, Japan.}

\begin{abstract}
Neutral atoms are among the leading platforms toward realizing fault-tolerant quantum computation (FTQC).
However, scaling up a single neutral-atom device beyond $\sim 10^4$ atoms to meet the demands of FTQC for practical applications remains a challenge.
To overcome this challenge, we clarify the criteria and technological requirements for further scaling based on multiple neutral atom quantum processing units (QPUs) connected through photonic networking links.
Our quantitative analysis shows that nanofiber optical cavities have the potential as an efficient atom-photon interface to enable fast entanglement generation between atoms in distinct neutral-atom modules, allowing multiple neutral-atom QPUs to operate cooperatively without sacrificing computational speed.
Using state-of-the-art millimeter-scale nanofiber cavities with the finesse of thousands, over a hundred atoms can be coupled to the cavity mode with an optical tweezer array, with expected single-atom cooperativity exceeding 100 for telecom-band transition of ytterbium atoms.
This enables efficient time-multiplexed entanglement generation with a predicted Bell pair generation rate of 100 kHz while maintaining a small footprint for channel multiplexing.
These proposals and results indicate a promising pathway for building large-scale multiprocessor fault-tolerant quantum computers using neutral atoms, nanofiber optical cavities, and fiber-optic networks.
\end{abstract}

\maketitle

\section{Introduction}
A major milestone in quantum technologies includes the construction of a large-scale quantum computer with promising applications in various areas such as machine learning~\cite{Rebentrost2014,kerenidis2016,Zhao2019,Yamasaki2020,Liu2021,yamasaki2022,doi:10.1126/science.abn7293,yamasaki2023quantum,yamasaki2023advantage}, quantum chemistry~\cite{Bauer2020,Lee2021,Reiher2017,Kim2022}, and cryptanalysis~\cite{Shor1997,gheorghiu2019,Gidney2021}.
A prominent example is Shor's algorithm for factoring 2048-bit integers, which requires two-qubit gate error rates reaching the $10^{-10}$ level~\cite{Gidney2021}.
To attain such a small error rate, it is necessary to replace qubits in the original quantum circuit with logical qubits encoded in a quantum error-correcting code to perform fault-tolerant quantum computation (FTQC). 
In FTQC, the computation is implemented as a sequence of logical operations acting on the logical qubits, allowing arbitrarily small logical error rates by scaling the code as long as the physical error rates are below a certain constant threshold.

Significant experimental progress has been made in recent years towards this goal among multiple platforms.
Here, we focus on neutral-atom-based quantum technologies, featuring the physical two-qubit gate error rate below $10^{-2}$ \cite{Evered2023}, which is expected to reach $10^{-3}$ by improving the laser systems and the excitation protocols~\cite{Evered2023,Jiang2023}.
Moreover, the number of trapped and controllable qubits has increased by orders of magnitude to the level of multiple thousands~\cite{Norcia2023,Manetsch2024}.
Logical quantum processing with the execution of small-scale quantum algorithms on logical qubits has also been demonstrated~\cite{Bluvstein2024}.
A particular merit of atom arrays is their flexible reconfigurability for two-qubit gate operation between nearly arbitrary pairs of atoms.
This is crucial for implementing high-rate quantum error-correcting codes, such as concatenated quantum Hamming codes~\cite{Yamasaki2024,Yoshida2024} and quantum low-density parity-check (LDPC) codes~\cite{Leverrier2015,Fawzi2018,Bravyi2024,panteleev2022asymptoticallygoodquantumlocally,leverrier2022,dinur2022goodquantumldpccodes}.
Despite these desired properties, the number of physical qubits in a single neutral-atom module is estimated to be limited to an order of $10^4$.
The typical restraints are the size of the field of view of objective lenses and the laser power to trap and control atoms, which cannot be increased indefinitely~\cite{Norcia2023,Manetsch2024}.
With this limit, problematically, we can encode only a few tens or hundreds of logical qubits in a single module while satisfying the logical two-qubit gate error rate of $10^{-10}$, even if we use the high-rate quantum error-correcting codes~\cite{Yoshida2024}.
This limitation poses a formidable challenge in scaling up a neutral-atom quantum computer to the regime of various useful applications.

The solution to this scalability issue involves a modular architecture that connects multiple modules with photonic links~\cite{Kimble2008, Covey2023, Wehner2018, Young2022, Monroe2014}.
In this architecture, intermodule operations are performed by quantum teleportation of states~\cite{Bennett1993} and gate teleportation~\cite{gottesman1999demonstrating,PhysRevA.62.052316} at the level of logical qubits.
For a single intermodule logical operation using teleportation,
we need heralded preparation of a large number of high-fidelity physical atom-atom Bell pairs on the order of hundreds or thousands, for the modules with about $10^4$ physical qubits.
To achieve even moderate connectivities across modules, such a large number of Bell pairs need to be produced on a timescale comparable to the error-correction cycle times.

\begin{figure*}[t]
    \centering
    \includegraphics[width=7.0in]{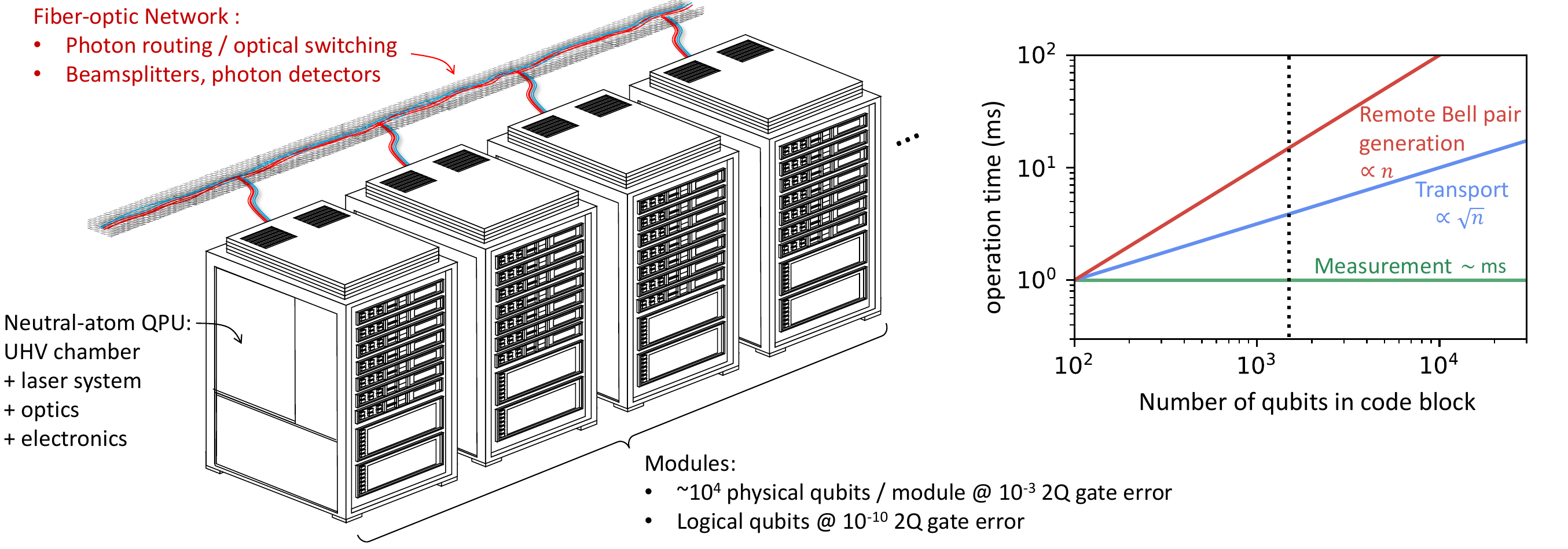}
    \caption{A multiprocessor fault-tolerant quantum computer realized by neutral atoms and fiber-optic network. 
    (Left) We envision neutral-atom quantum processing units (QPUs) connected with an optical fiber network to enable fault-tolerant inter-module operations. 
    (Right) Time budgeting of networked neutral-atom QPUs\@. 
    Typically, in a practical regime of the logical error rate $10^{-10}$, we need hundreds, or beyond a thousand, physical qubits in a single code block~\cite{Bravyi2024,Yoshida2024,Xu2024} (vertical dotted line), which encodes one to tens of logical qubits depending on the choice of code.
    Within a single neutral-atom module, the time cost of atom transport for quantum error correction (blue) typically scales as $\propto \sqrt{n}$ or better, where $n$ is the number of physical qubits in a code block for 2D atom arrays~\cite{Xu2024}. 
    In contrast, protocols for remote entanglement generation are inherently sequential, incurring a linear time cost $\propto n$ (red).
    A measurement can be parallelized over a large number of atoms, typically over a millisecond (green).
    The time of each gate in the module is not shown in the plot since the gates are performed in parallel with a much shorter timescale of $\SI{}{\micro \second}$.
    For this plot, we used $10^5$ Bell pairs per second, one of the fastest rates of proposed remote entanglement generation for the atom array system with the transport time cost taken into account~\cite{Li2024}.
    }
    \label{fig:vision}
\end{figure*}

Currently, the rate of light-mediated remote entanglement generation lags far behind the local computation speed, at a sequential generation of Bell pairs at hundreds of Hz \cite{Stephenson2020} and a projected rate of $\sim 100$ kHz \cite{Li2024}, while local gates already operate in parallel on a large number of qubits in microseconds \cite{Evered2023}.
Thus in scaling up the neutral-atom quantum computer while maintaining its flexibility, the photonic link is likely to become a significant bottleneck.

In this paper, we analyze and discuss potential pathways for scaling up neutral-atom quantum computers beyond a single module (Fig.~\ref{fig:vision}) from the perspectives of hardware development and fault-tolerant protocol design.
Sec.~\ref{sec:criteria} addresses the requirements of networked modules from an FTQC standpoint.
In Sec.~\ref{sec:neutral_atom}, we discuss concrete physical operations on a neutral atom platform to identify requirements for networking.
Sec.~\ref{sec:network} discusses time-multiplexed remote entanglement generation protocols for networking and we analyze their requirements for further scaling; 
to meet the requirements, we propose a prospective implementation using nanofiber optical cavities, which have the potential to scale up by time, wavelength, and channel multiplexing. 
Finally, in Sec.~\ref{sec:conclusion}, we conclude with an outlook for hardware and fault-tolerant architecture development.

\section{Renewed criteria for physical implementation of quantum computers}
\label{sec:criteria}

Quantum computation is conventionally represented by a quantum circuit designed to solve a computational problem; 
however, executing the original quantum circuit directly on a noisy quantum device during experiments does not yield a correct solution.
FTQC provides a technique to suppress the effect of noise while using noisy quantum devices at a non-zero physical error rate~\cite{gottesman2010};
instead of the original circuit, a fault-tolerant circuit can be executed on the physical qubits, which simulates the original circuit using error-suppressed logical qubits encoded in a quantum error-correcting code of many physical qubits.
As the size of the problem to be solved increases, the required code size for the fault-tolerant protocol, i.e., the number of physical qubits per code block, also increases to attain sufficient suppression of logical error rates.

Most of the current efforts towards FTQC focus on single-module improvement following DiVincenzo's five criteria for quantum computation~\cite{DiVincenzo2000}, namely,
(i) a scalable physical system with well-characterized qubits; 
(ii) the ability to initialize the state of the qubits to a simple fiducial state, such as $\ket{000\cdots 0}$; 
(iii) long relevant decoherence times, much longer than the gate operation time; 
(iv) a universal set of quantum gates; and 
(v) a qubit-specific measurement capability.
Since the introduction of DiVincenzo's criteria, substantial technological advances have been made, leading to a more precise understanding of the challenges in building quantum computers.
One of the challenges is the requirement of scalability: the number of qubits in the system should scale up to millions for some practical applications~\cite{Gidney2021} while maintaining a precise calibration.
However, the number of qubits that can be incorporated within a single module turns out to be constrained by several technical limitations in the physical implementation of most qubit systems. 
For example, in the case of neutral atoms in an optical tweezer array, the number of atoms in a single chamber is technically limited by factors including available laser power and the field of view of the objective lens, making it challenging to realize scalable FTQC through a single standalone module.

In view of the crucial role in addressing scaling challenges,
we propose criteria to interconnect multiple quantum modules to build a multiprocessor fault-tolerant quantum computer, as illustrated in Fig.~\ref{fig:vision}.
Instead of arbitrarily scaling up a single module, each module should only satisfy a finite set of technological requirements necessary for its function and quality, for example, the number of physical qubits and the physical error rates.
Using such modules, a large-scale quantum computer can be designed in a modular way, comprising multiple independent modules that can be flexibly replaced and recombined.
For scaling up, variations in module qualities should be allowed, but the modules should demonstrate a threshold behavior, ensuring that the complexity of building the quantum computer remains manageable as long as the average quality of the modules exceeds a certain threshold, for example, given by the threshold of a fault-tolerant protocol.
To embody principles of finite technological requirement, modularity, and threshold behavior, 
it is natural to integrate DiVincenzo's two additional criteria for quantum communication into the quantum computer design ~\cite{DiVincenzo2000}:  the ability to (vi) interconvert stationary and flying qubits, and (vii) faithfully transmit flying qubits between specified locations.

Scalable physical implementation of a multiprocessor fault-tolerant quantum computer thus calls for the following modified criteria:
\begin{itemize}
    \item (i) Fixed-size module with a finite number of qubits;
    \item  (ii) Initialization operation to set the state of the qubits within each module into a fixed state, such as $\ket{000\cdots}$, with a finite error rate below a threshold constant;
    \item (iii) Gate operations on the qubits in each module with a finite error rate below a threshold constant; 
    \item  (iv) Universal gate set composed of a finite number of gates;
    \item  (v) Measurement operation on the qubits in each module with a finite error rate below a threshold constant;
    \item  (vi) The interface of each module to be linked with other modules, with a finite error rate below a threshold constant;
    \item  (vii) The link connecting the interfaces of different modules to transmit quantum states within, at most, a fixed distance, with a finite error rate below a threshold constant.
\end{itemize}

Recent advances in quantum technologies have made substantial progress in fulfilling criteria~(i)--(v) as in conventional DiVincenzo's criteria for quantum computers at a fixed scale.
However, the importance of criteria~(vi) and~(vii) for modularity has not been thoroughly considered from a combined standpoint of experimental implementation of FTQC\@.
The remainder of this paper discusses the pathways to satisfy criteria (vi)-(vii) in a scalable manner, focusing on the hardware development of quantum processing units (QPUs) realized by neutral atoms.

\begin{figure*}[t]
    \centering
    \includegraphics[width=7.0in]{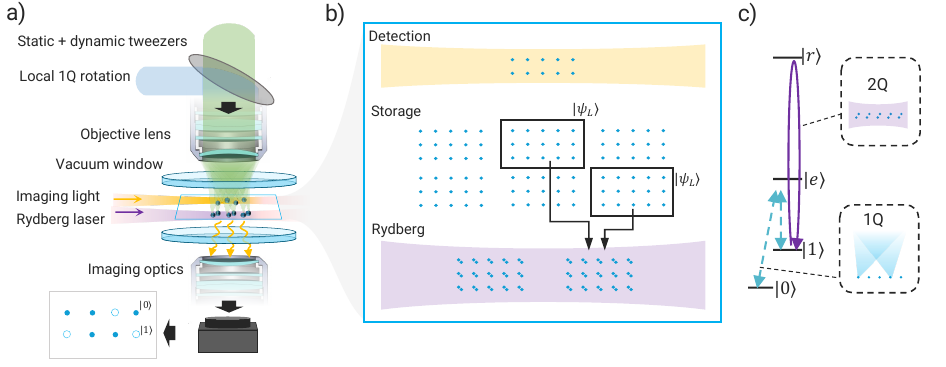}
    \caption{Overview of a module of neutral-atom quantum processing unit (QPU). 
    (a) Typical experimental apparatus comprising an ultra-high-vacuum (UHV) chamber with vacuum windows for optical access, having a high-numerical-aperture (high-NA) objective lens focused on a plane where atoms are trapped.
    Static and dynamic tweezer beams are generated by spatial light modulators (SLMs) and acousto-optic deflectors (AODs), focused with the lens for tight trapping.
    Dichroic mirrors are used to combine lights at several wavelengths to be sent through the objective lens for individually addressed qubit control, whereas global or quasi-local addressing light (Rydberg laser and imaging light) can be sent along the two-dimensional plane for zoned architecture.
    Another objective lens can be used to collect fluorescence light to identify qubit states.
    (b) Zoned architecture with local Rydberg and imaging lasers for spatially selective imaging and Rydberg gates can be used to perform parallel gate operations and measurements interleaved by qubit shuttling while maintaining a long coherence time for idle qubits.
    (c) Single and two-qubit gate operations for neutral atoms are illustrated. 
    }
    \label{fig:qpu_overview}
\end{figure*}

\section{Neutral-Atom QPU for FTQC}
\label{sec:neutral_atom}

Reconfigurable atom arrays are promising platforms for quantum information processing, due to their long coherence time along with long-range, high-fidelity, and parallel gate operations~\cite{Bluvstein2024}.
In building a fault-tolerant quantum computer, the threshold in the fault-tolerant protocols determines the requirements for finite precision in criteria~(i)--(vii).
For state-of-the-art fault-tolerant protocols, the threshold values are typically of the order $10^{-2}$ -- $10^{-3}$ under a conventional error model~\cite{Yoshida2024,Knill_2005,PhysRevA.68.042322,10.5555/2011814.2011815,Vuillot_2019}.
Furthermore, the following physical operations are required at the module level for criteria~(i)--(v):
$\ket{0}$-state preparation, $Z$-basis measurement, Pauli gates ($X$, $Y$, and $Z$), Clifford gates ($H$, $S$, and \textsc{CNOT}) and non-Clifford gates ($T$)~\cite{gottesman2010}. 

In Sec.~\ref{sec:neutral-atom-qpu}, we outline how these physical operations are typically realized in neutral-atom qubit systems.
Furthermore, we discuss the current and projected future performances for implementing FTQC with neutral atoms in Sec.~\ref{sec:ftqc_neutral_atom}.

\subsection{Neutral-atom QPU}
\label{sec:neutral-atom-qpu}

A typical neutral-atom qubit system is illustrated in Fig.~\ref{fig:qpu_overview}(a). 
Individual atoms are trapped in optical tweezers and qubits are encoded in their internal degrees of freedom \cite{Bluvstein2022,Ma2022,Jenkins2022,Barnes2022}, with coherence times ranging from seconds to tens of seconds. 
The trapped atoms can be moved among separate zones to perform different types of operations as shown in Fig.~\ref{fig:qpu_overview}(b).

\subsubsection{Atom loading, state preparation, and measurements}

In a neutral-atom system, room-temperature atomic vapor is initially laser-cooled into a magneto-optical trap (MOT), from which individual atoms are loaded into an array of tightly focused optical tweezers~\cite{Nogrette2014}.
Collisional blockade and sorting with dynamic tweezers realize a low-entropy array of single atoms, operated by acousto-optic deflectors (AODs)~\cite{Barredo2016,Endres2016}.
Then the qubit states are initialized (for example, in $|00\cdots0\rangle$) by optical pumping with fidelity reaching 99.8~\%~\cite{Ma2022,Evered2023}.

One of the major operational drawbacks of neutral-atom qubits is the loss of atoms from the trap, induced by collisions with background molecules in the chamber~\cite{2021Schymik} and heating during imaging~\cite{Chow2023,Lis2023}, among other reasons~\cite{Ma2023,Scholl2023}.
Several techniques are being developed to mitigate these problems, such as continuous reloading based on a dual-species platform~\cite{Singh2023,Singh2022} and the use of metastable states of alkaline-earth atoms~\cite{Gyger2024,Norcia2024}.
In such schemes, newly loaded atomic qubits are prepared away from the computing zone and then transported to replace old trap sites, which represents an important milestone toward continuous operation beyond the trap lifetime of the atoms.
These techniques can be incorporated into fault-tolerant quantum computing for quantum teleportation of states~\cite{Nolleke2013} and Knill teleportation-based quantum error correction~\cite{Knill_2005}.
Atom losses can be detected by imaging, allowing the identification of leakage error locations that convert leakage errors into erasure errors with known locations, thereby improving the decoding performance in quantum error correction~\cite{Sahay2023}.
The erasure conversion is also possible for other leakage-inducing operations of atomic qubits, such as the excitation to Rydberg states~\cite{Wu2022,Ma2023,Scholl2023}.

State-selective measurements, i.e., $Z$-basis measurements of qubits, are performed in parallel by collecting the fluorescence signal from one of the qubit states of atoms, with a typical timescale of a millisecond or less, to achieve high detection fidelity reaching 99.9\% ~\cite{Chow2023,Manetsch2024} with demonstration of small atom loss probability below 1~\%~\cite{Manetsch2024}.
An alternative method for qubit measurements is via optical cavities, which, albeit sequential, has a fast measurement time of tens of microseconds; 99.5~\% measurement fidelity and percent-level atom loss within $\SI{50}{\micro \second}$ measurement duration have been demonstrated in Refs.~\cite{Tamara2021, Deist2022}. 

\subsubsection{Gate operations} 

Single- and two-qubit gate operations for neutral atoms are illustrated in Fig.~\ref{fig:qpu_overview}(c).
Fast single-qubit $X$ and $Z$ rotations are possible by Raman coupling~\cite{Ma2022, Jenkins2022, Levine2022}, with fidelity exceeding 99.9~\% at a gate speed of a microsecond, thereby realizing arbitrary single-qubit rotations including Pauli and Clifford gates.
The rotations can be performed globally using large laser beams that cover the entire or part of the atom array, or individually by applying tightly focused control lasers.

Two-qubit controlled phase gates are realized using the Rydberg blockade effect~\cite{Saffman2010} by exciting atoms from one of the qubit states to highly excited Rydberg states within a typical timescale of a microsecond.
The fidelity exceeds 99~\%~\cite{Evered2023,Paper2024} and expected to reach 99.9~\% with better excitation protocols and improved laser systems~\cite{Evered2023, Jiang2023, Jandura2023, Bluvstein2022}.
As the Rydberg blockade occurs at a distance of typically micrometer order, programmable and arbitrary pairwise interactions can be designed via an atom rearrangement sequence such that pairs of atoms to interact are placed within a certain radius when a Rydberg excitation laser is applied~\cite{Bluvstein2022}.

\begin{figure*}[t]
    \centering
    \includegraphics[width=7in]{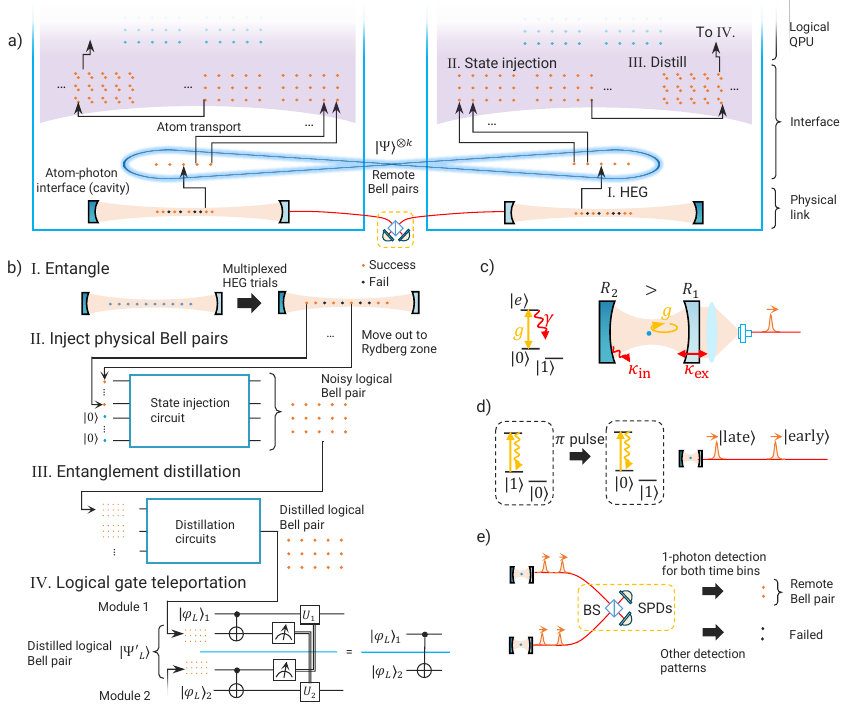}
    \caption{Optical link between neutral-atom QPU modules. 
    (a) Atom-photon interfaces, such as optical cavities, can be integrated into neutral-atom QPU modules by placing them on the focal plane of the objective lens.
    Through the interface, atoms in different modules can be entangled into Bell pairs to be used in subsequent logical operations.
    (b) The generated physical Bell pairs can be transported into a local computing zone (I), to be encoded into noisy logical Bell pairs by state injection (II), followed by entanglement distillation using logical operations on the code blocks to reach the desired error rate (III).
    These logical Bell pairs are then used for remote logical-qubit operations, such as quantum teleportation of states or gate teleportations (see text).
    In IV, we illustrate the logical gate teleportation between arbitrary logical code blocks $\ket{\varphi_L}_1$ and $\ket{\varphi_L}_2$ in separate modules.
    (c) Atom-photon coupling with three-level atoms.
    The transition between one of the qubit states and the excited state $|e\rangle$ is resonant with the cavity at a coupling rate $g$. 
    The cavity is also characterized by the internal loss rate $\kappa_{\mathrm{in}}$, the coupling to the external mode $\kappa_{\mathrm{ex}}$, and the atom state decay $\gamma$.
    (d) Atom-photon entanglement is created by emitting a photon from the atoms, for example, with photons in the two time bins that are correlated with the qubit state.
    (e) A typical setup for heralded entanglement generation (HEG) involves atom-photon interface, an optical link, and a detection module comprising nonpolarizing beamsplitter (BS) and single-photon detectors (SPDs).
    The photons are measured after being interfered at a beamsplitter and the successful trials are characterized by a single detector click for each time bin, which projects the qubit pair to the maximally entangled (Bell) state~\cite{Barrett2005}.
    }
    \label{fig:interface}
\end{figure*}

\subsection{FTQC with neutral atoms and quantum-error-correction cycle time}
\label{sec:ftqc_neutral_atom}

The ability to move atom qubits for implementing long-range and parallelized interactions makes the neutral-atom platform a particularly promising candidate for implementing constant-space-overhead fault-tolerant protocols using quantum error-correcting codes with high rates of logical qubits per physical qubit~\cite{Gottesman2014,Fawzi2018,Yamasaki2024,Yoshida2024}.
Conventional fault-tolerant protocols, such as those using surface code~\cite{10.1063/1.1499754} and concatenated Steane code~\cite{10.5555/2011665.2011666}, incur a polylogarithmically growing overhead of the number of physical qubits per logical qubit.
Compared to conventional protocols, constant-space-overhead protocols are expected to significantly reduce the overhead of FTQC ~\cite{Yamasaki2024,Yoshida2024,PhysRevLett.129.050504,Xu2024,Bravyi2024}.

Several approaches are known for constructing constant-space-overhead fault-tolerant protocols~\cite{Gottesman2014,Fawzi2018,Yamasaki2024,Yoshida2024}.
Among them, the use of high-rate quantum concatenated codes~\cite{Yoshida2024,Yamasaki2024} is particularly advantageous for atomic qubit platforms, where finite-size codes are employed at each level of concatenation, and error suppression to the desired logical error rate is achieved by increasing the concatenation level.
A characteristic feature of the code-concatenation approach is that the finite-size quantum error-correcting code used in each concatenation inherently provides an abstraction layer for modularity. 
Once the code block at a fixed concatenation level is realized as a module, multiple modules can be combined to implement another finite-size code to be concatenated at the next concatenated level for further error suppression~\cite{Yoshida2024,Yamasaki2024}.
A high-threshold scheme with high-rate concatenated codes has been proposed in Ref.~\cite{Yoshida2024}, where the $C4/C6$ architecture originally proposed in Ref.~\cite{Knill_2005} is used at the physical level and constant space overhead is achieved asymptotically by switching the codes to quantum Hamming codes~\cite{Yamasaki2024}.
Another approach to a constant-space-overhead FTQC is based on quantum LDPC codes~\cite{Gottesman2014,Fawzi2018}.
In this approach, error suppression to the desired logical error rate is achieved by increasing the code size of a family of quantum LDPC codes.
A hardware-efficient scheme for performing FTQC with high-rate quantum LDPC codes on reconfigurable atom arrays has been proposed in Ref.~\cite{Xu2024}.
Both approaches have the potential to be realized on the neutral-atom platforms.

The spatially multiplexed optical control of physical qubits using the AOD-generated optical tweezer array and control laser beams permits a natural abstraction of logical qubits control for quantum error-correcting codes of neutral-atom physical qubits, without increasing the control lines.
This architecture leverages the programmability and bandwidth of optical control, along with the transversal operations of quantum error-correcting codes~\cite{Bluvstein2024}.
For example, atoms in the same code block can be moved together and transversal single- and two-qubit gates can be performed by simultaneously illuminating control laser beams to the entire code block, as illustrated in Fig.~\ref{fig:qpu_overview}(b).

A single cycle of mid-circuit measurement and feedforward operations is expected to take on the order of a millisecond, which is dominated by the measurement time.
Combined with the atom rearrangement time~\cite{Bluvstein2022,Xu2024}, a single cycle of quantum error correction (QEC) and logical gate operations is expected to be of the order of milliseconds.
In this case, the time overhead per logical gate operation is expected to scale as $\sim \sqrt{n}$ for large $n$, due to the transport time of the two-dimensional atom array~\cite{Xu2024}, likely to determine the module-level clock speed.

\section{Photonic Networking of Neutral-Atom QPUs}
\label{sec:network}

While the number of physical qubits in a single neutral-atom module is restricted by several fundamental limitations, the number of physical qubits needs to scale as the requirements for logical qubit number and error rate become more stringent.
Analogously to the multiprocessor system architecture widely used in modern classical computers, the fundamental requirement naturally motivates the introduction of multiple neutral-atom quantum processors as modules interconnected by photonic links, as captured by our modified criteria (vi) and (vii).

In this section, we discuss the requirements and implementations of networked neutral-atom QPUs\@.
First, Sec.~\ref{sec:link_ftqc} introduces methods for linking multiple logical neutral-atom QPUs using Bell pairs prepared across different neutral-atom modules through the photonic links.
In Sec.~\ref{sec:heg}, we focus on the concrete implementation of neutral-atom systems and describe a standard photon-assisted protocol to generate a remote physical Bell pair between atoms in different modules.
In Sec.~\ref{sec:fast_heg}, we describe an efficient time-multiplexed protocol to fully utilize the optical channel in the presence of slow operations, such as atom transport and qubit initialization, to identify the requirement for a scalable atom-photon interface.
In Sec.~\ref{sec:nanophotonics}, the advantages and challenges of waveguide cavities are discussed, considering the aforementioned requirements.
To overcome the limitations, in Sec.~\ref{sec:nanofiber}, we propose the use of nanofiber optical cavities to develop a scalable atom-photon interface for realizing a multiprocessor fault-tolerant quantum computer.

\begin{figure*}[t]
    \centering
    \includegraphics[width=7.0in]{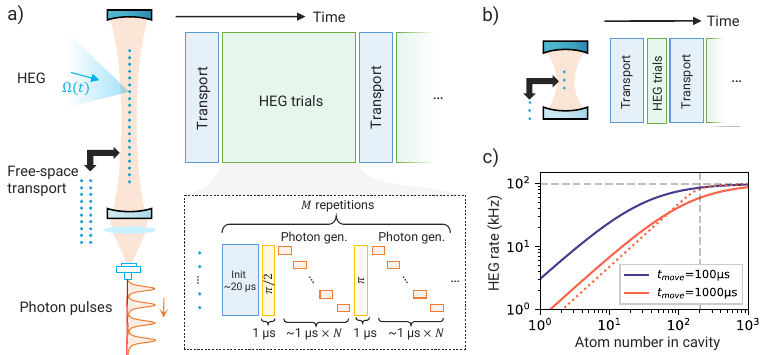}
    \caption{Time-multiplexed entanglement generation.
    (a) With a large number of atoms in the cavity, slow operations, including atom transport and initialization, can be performed in parallel, and the time cost is spread over many sequential photon generation trials.
    Photon generation for HEG can be performed with high fidelity by applying a focused excitation laser to the target atom while inducing a light shift to the other atoms with addressed lasers, thereby uncoupling from the cavity.
    (b) In contrast, for a smaller number of atoms, the operation is dominated by atom transport and initialization times, while the photonic channel is mostly left idle.
    (c) Entanglement generation (success) rate for varying numbers of atoms transported in parallel.
    HEG success probability $p_e^2/2 = 0.2$, pulse separation $\Delta t= 1\ \mu$s, and initialization time are $\SI{20}{\micro \second}$, and considered fast ($\SI{100}{\micro \second}$, purple) and slow ($\SI{100 0}{\micro \second}$, orange) atom transport~\cite{supp}.
    The dashed horizontal line is the maximum possible rate for no transport and initialization cost, $p_e^2/4\Delta t = $100 kHz.
    With two hundred atoms in the cavity (vertical dashed line), the rates are nearly saturated at 90\% of $p_e^2/4\Delta t$ for the fast-transport case and 65\% for the slow-transport case.
    Zoned multiplexing described in the Supplemental Materials~\cite{supp} improves the rate for the slow-transport case (dotted line) by parallelizing atom transport and photon generation to reach nearly 90\% of the maximum possible rate for two hundred atoms.
    }
    \label{fig:time-mux}
\end{figure*}

\subsection{Roles of photonic links in FTQC}\label{sec:link_ftqc}

The aim of introducing photonic links connecting the neutral-atom modules is to prepare logical Bell states between code blocks in different neutral-atom modules, as shown in Fig.~\ref{fig:interface}(a).
Logical Bell states, with their logical error rates sufficiently suppressed, can be used for quantum teleportation of logical states and gate teleportation between modules.
As will be described later in this section, the photonic link is used to generate a physical Bell state between atoms in different modules, at a physical error rate of $\sim 10^{-2}$ while maintaining high entanglement generation rates~\cite{Li2024, Saha2024}\footnote{While there exist proposals to achieve remote Bell pair infidelity of $10^{-3}$, narrow detection time-windowing or postselection are required \cite{Li2024, Saha2024}, in addition to exquisite optical alignments and preparation of high-fidelity photon detector. Furthermore, the overall gate teleportation accumulate infidelity from local gates and measurements in addition to the Bell pair infidelity, thus achieving high-fidelity remote physical CNOT gates are challenging. Instead, here we consider the case of a slightly relaxed Bell pair infidelity condition on the order of $10^{-2}$ and take advantage of rapid error reduction in logical-level entanglement distillations to seek a more realistic approach.}.
A key challenge to be addressed here is determining how to use the noisy atom-atom Bell states prepared at a physical error rate of up to $\sim 10^{-2}$ for FTQC\@.

A straightforward method of using the noisy inter-module atom-atom physical Bell state would be to directly use for gate teleportation to apply a noisy physical \textsc{CNOT} gate between the modules at the physical error rate again of order $\sim 10^{-2}$; 
since the thresholds for fault-tolerant protocols are typically on the order of $10^{-3}$ or $10^{-2}$~\cite{Yoshida2024,Knill_2005,PhysRevA.68.042322,10.5555/2011814.2011815,Vuillot_2019}, such a noisy CNOT gate may be insufficient for achieving FTQC. 
Even if the physical error rates marginally surpass the threshold, the fault-tolerant protocol may require an impractically large overhead, necessitating a sufficient margin between the threshold and the physical error rate.

To achieve tractable overhead in FTQC, noisy Bell states should be transformed into ones with their errors sufficiently suppressed for FTQC using entanglement distillation protocols~\cite{Deutsch1996, Bennett1996_mixedstate,Bennett1996_Purification,RyutarohMatsumoto_2003,Devetak2005,Rozpdek2018,Krastanov2019}.
However, implementing these entanglement distillation protocols directly with physical operations on neutral atoms in each module may be insufficient\@; 
if we proceed in this manner, errors from physical operations typically of the order of $10^{-3}$, as discussed in Sec.~\ref{sec:neutral_atom}, will accumulate during the protocol. 
Consequently, the resulting Bell state from the protocol may fail to surpass the threshold error rate by an ample margin.

A promising approach to overcome this challenge involves combining state injection and entanglement distillation implemented by logical operations on the code blocks in each module, as shown in Fig.~\ref{fig:interface}(b).
State injection is a technique for encoding physical qubit states into logical states of a quantum error-correcting code~\cite{knill1996concatenated,Knill_2005,10.1063/1.1499754,PhysRevA.80.052312,Litinski2019magicstate}.
Applying the state injection to the noisy atom-atom physical Bell states across the modules, the modules can share the noisy logical Bell states encoded in the code blocks of the modules.
The logical Bell state prepared by state injection has a logical error rate of order similar to that of the physical error rate of the initial physical Bell state~\cite{9761242,lodyga2015simple}, where the errors from the physical gates are not dominant because the initial physical Bell states prepared over the modules are noisier than the local gate operations in the modules.
Entanglement distillation, implemented by local logical operations, transforms noisy logical Bell states into logical Bell states at an arbitrarily small logical error rate.
As the errors of the logical operations are suppressed, the threshold of entanglement distillation is much better than that of fault-tolerant protocols implemented by noisy physical operations.

To estimate the required speed for physical Bell pair generation between modules for FTQC, the overhead of entanglement distillation, that is, the required number of physical Bell pairs (and the run-time) for preparing logical Bell states at a target logical error rate by entanglement distillation should be considered.
The estimation of such an overhead largely depends on the details of the quantum error-correcting codes used in fault-tolerant protocols and the entanglement distillation protocols. 
In the following analysis, we analyze a typical regime in which, for modules composed of code blocks of $n$ physical qubits, generating $\sim n$ physical Bell states per QEC cycle is sufficient to implement FTQC having multiple modules without sacrificing computational speed.

\subsection{Heralded entanglement generation (HEG)}
\label{sec:heg}

The generation of remote atom-atom Bell pairs over different neutral-atom modules forms the basis for realizing qubit operations across modules, such as quantum teleportation of states~\cite{Olmschenk2009}, gate teleportation~\cite{Goto2009}, and remote parity readout~\cite{Nickerson2014}.
An exemplary protocol for remote atom-atom entanglement generation involves the single-photon emission of the atom-state-dependent photonic state, followed by the measurement of emitted photons.
Here, we focus on a protocol based on time-bin encoded photons \cite{Barrett2005}, and a similar discussion holds for polarization or frequency-based photon encodings \cite{Duan2006,Moehring2007}.
We consider each atom to be placed in a cavity for high-efficiency photon collection \cite{Young2022}, where the internal state of the atom is in a superposition of qubit states $(|0\rangle+|1\rangle)/\sqrt{2}$ (Fig.~\ref{fig:interface}(c)).
Sequential photon emission effectively results in an atom-photon entangled state$(|0,\mathrm{early}\rangle+|1,\mathrm{late}\rangle)/\sqrt{2}$, where $\{\ket{\mathrm{early}},\ket{\mathrm{late}}\}$ represents a time-bin encoding, as illustrated in Fig.~\ref{fig:interface}(d).
The photon pulses from separate atom qubits are interfered with a 50-50 beamsplitter, and photon-detection signals from single-photon detectors (SPDs) are recorded for each time bin (early and late), as shown in Fig.~\ref{fig:interface}(e).
The desired photon detection pattern (only a single photon detection for each early and late time bin) results in the projection of a remote atom-atom pair onto a Bell state~\cite{Barrett2005}.
The protocol is known as the heralded entanglement generation (HEG).

Following successful HEG, atoms are transported out of the cavity using dynamic tweezers and buffered in free space until a sufficient number of Bell pairs are ready for state injection and distillation, as described in Sec.~\ref{sec:link_ftqc}. 
Such remote logical operations impose strict requirements for the Bell pair generation rate, necessitating the preparation of a large number of physical Bell pairs for each remote logical gate operation scaling linearly with the number of logical qubits of code blocks.
Even for relatively modest remote connectivity of a single-block remote two-qubit gate associated with each module in every few QEC cycles, the requirement is stringent to achieve error rates for notable applications, e.g., of order $10^{-10}$ (Fig.~\ref{fig:vision}).
For example, with a projected clock time of below one millisecond~\cite{Bluvstein2024}, tens of logical qubits to be injected in a code block \cite{Xu2024, Yoshida2024} and an overhead for distillation at a few tens \cite{Krastanov2019}, then MHz entanglement generation would be required so as not to become a significant bottleneck of computation.
This requirement will become even more demanding as the module-level operations of the neutral-atom logical qubit platform scale up and speed up.

The fidelity of physical entangled atom pairs is also a crucial metric that should exceed the threshold constant determined by a fault-tolerant protocol for state injection and logical entanglement distillation.
In state-of-the-art implementations, remote Bell pair fidelities above 90\% have been demonstrated with neutral atoms and trapped ions~\cite{Stephenson2020, Ritter2012, Saha2024}.
Various technical improvements, for example, better qubit control, suppressed loss in optical elements, and the adoption of windowed herald strategies are expected to allow for sufficiently high fidelity beyond 99\%~\cite{Li2024, Saha2024}.

\subsection{Fast HEG with atom array in a cavity}\label{sec:fast_heg}
High-rate entanglement generation requires high-efficiency photon-collection systems and a protocol to fully utilize photonic channels without significant idle times.
In this subsection, we introduce optical cavities as an efficient photon collection mechanism for fast HEG and discuss the efficient usage of these systems using a time-multiplexed protocol to maximize the use of photonic channels in the presence of slow auxiliary operations.

Optical cavities, comprising a pair of mirrors surrounding the atom, enhance the atom-photon coupling by the circulation of light \cite{Reiserer2015,Reiserer2022}, as shown in Fig.~\ref{fig:time-mux}(a). 
They provide a significant improvement in photon collection efficiency compared to free-space optics, by enhancing emission into the cavity mode~\cite{Young2022}.
In particular, single-photon emission with the desired temporal mode can be induced while maintaining a high total emission probability using an appropriate temporal profile of the excitation laser beam~\cite{Vasilev2010,Gorshkov2009,Utsugi2022,supp}.
The photon then exits the cavity through the mirror with a lower reflectivity $R_1$ for fiber-optic photon routing (Fig.~\ref{fig:time-mux}).
There is a fundamental trade-off between the probability of photon emission $p_e$ and the duration of the photon pulse $\tau$, limiting the HEG success rate \cite{Goto2019}.
In the case of a simple Gaussian wave packet, the HEG rate takes an optimum at $\kappa_{\mathrm{ex}} \sim \kappa_{\mathrm{in}} + g$, yielding $p_e^2 / 2\tau > $ MHz order for state-of-the-art cavities,
for example, with single-atom internal cooperativity of $C_{\mathrm{in}} = g^2/(2\gamma \kappa_{\mathrm{in}}) \gtrsim 100$ (see Supplemental Material for more details~\cite{supp} and Ref.~\cite{Goto2019}).
With sufficient separation of the pulses, e.g., $\Delta t = 10 \times \tau$ to reduce the overlap to the level below $10^{-5}$, the predicted rate is hundreds of kilohertz\@.

While the aforementioned consideration provides a reasonable HEG rate, in practice, auxiliary operations such as atom transport and initialization incur significant time costs, on the order of 100 $\mu$s in total, thereby dominating the cycle time for a conventional situation with a single atom (or a few atoms) inside a cavity.
Mitigating this effect requires parallel transport and the preparation of a large array of atoms to minimize channel downtime by time multiplexing~\cite{Huie2021,Li2024}, as illustrated in Fig.~\ref{fig:time-mux}(a) and~(b). 
For an atom array with a large number $N$ of atoms transported and initialized in parallel, the dominant time cost is incurred by the sequential photon generation from individual atoms in the ideal case discussed above (Fig.~\ref{fig:time-mux}(a)).
In contrast, for a small $N$, the transport time dominates the time cost, limiting the generation rate of atom-atom entangled pairs (Fig.~\ref{fig:time-mux}(b)).

As shown in Fig.~\ref{fig:time-mux}(c), the HEG rate increases linearly with the number of atoms and quickly saturates near the bound $p_{\mathrm{e}}^2/4\Delta t$ = 100 kHz at approximately $N \sim 200$.
Notably, saturation is observed at similar atom numbers even for a significantly slower transport time (as shown by the orange solid line in Fig.~\ref{fig:time-mux}(c)), with a difference between the two rates at 30\% for $N=200$.
This difference reduces further to less than 5~\% with zoned operations inside the cavity for a more optimal use of the channel by parallelizing atom transport and photon generation across the two sets of atom arrays (see Supplementary Material for details~\cite{supp}).

\begin{figure*}[t]
    \centering
    \includegraphics[width=7.0in]{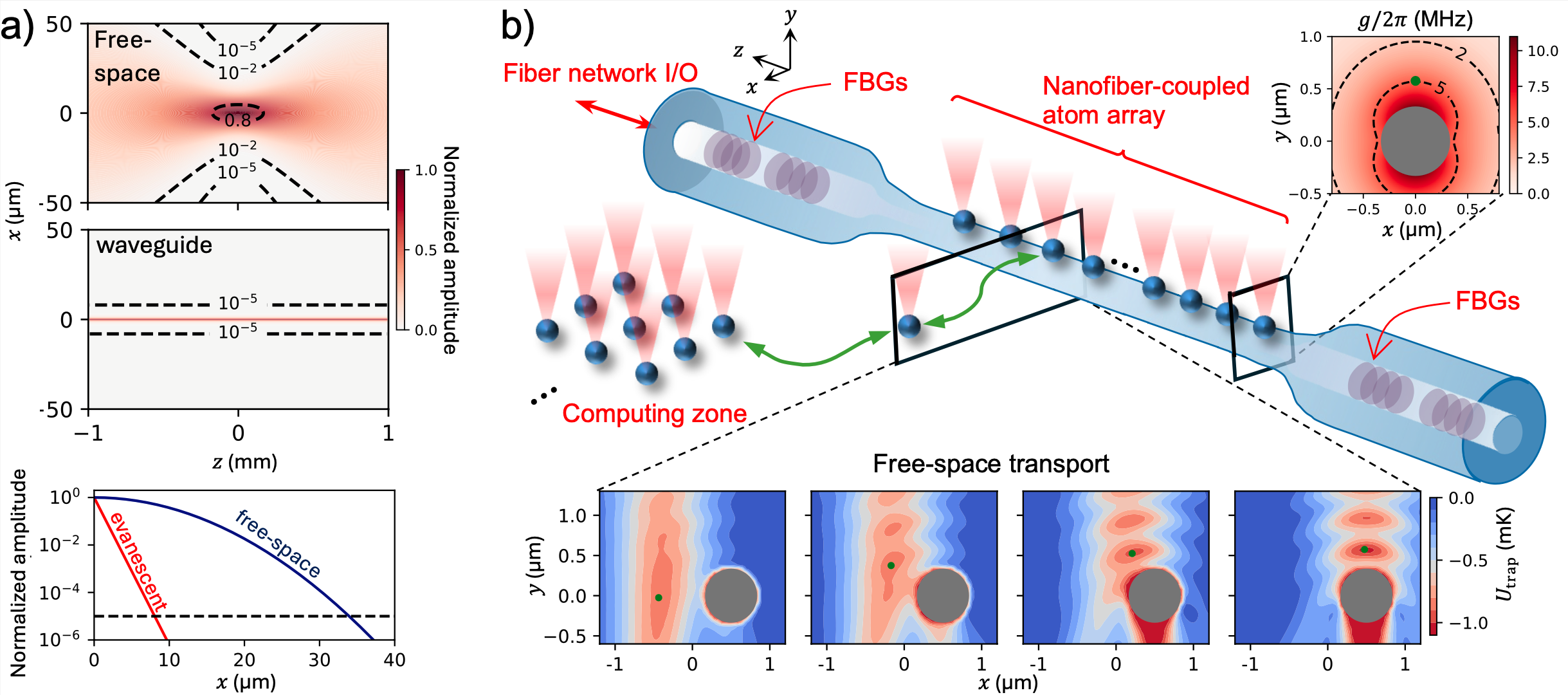}
    \caption{Nanofiber cavity quantum electrodynamics (QED)\@.
    (a) Comparison of cavity modes between the free-space and waveguide cavities.
    Waveguides are homogeneous in the cavity direction $z$ with a small mode area, whereas free-space cavities require a region with finite spread of the field amplitude distribution.
    The modulation of field amplitude along the $z$ (standing wave) is not shown for simplicity.
    The bottom figure compares the mode profile along the radial direction $x$ with $z=0$.
    Waveguide modes decay rapidly to allow sufficient decoupling with $<\SI{10}{\micro\meter}$.
    (b) The nanofiber cavity consists of a central nanofiber region where atoms are coupled, while the outer region has a larger diameter fiber (normal fiber or microfiber) where the FBGs are inscribed.
    The optical tweezer incident on the nanofiber creates a standing wave, which ensures a stable relative position of an atom to the fiber surface, stabilizing the coupling (bottom inset, colors indicating the simulated trap depth $U_{\mathrm{trap}}$).
    By moving the optical tweezer along the $x$ direction, atoms can be coupled in and out of the cavity mode at the nearest trap site (the green dot indicates the trap minimum).
    For the ytterbium atom in ${}^{3}P_0$ metastable state, with a state-insensitive tweezer trap at 759~nm, the atom-photon coupling rate $g$ is $\sim 2\pi\times$5~MHz, giving predicted single-atom internal cooperativity $C_{\mathrm{in}} > 100$ for the finesse of production-level nanofiber cavity~\cite{horikawa2024}.
    }
    \label{fig:nanofiber}
\end{figure*}

\subsection{Nanophotonic cavities for channel multiplexing}
\label{sec:nanophotonics}
The near-optimal channel usage for a given cavity quality and atomic transition is expected to be realized by the above considerations to provide entangled atom pairs at hundreds of kilohertz, potentially sufficient for small-scale FTQC units; 
however, the speed is fundamentally bounded by the number of photonic channels available because of sequential operations over the channel, making it challenging to scale up further.
Thus, as individual modules continue to improve both in qubit number and in operation speed, remote entanglement generation becomes a significant bottleneck in performing even moderately distributed operations in a multi-module architecture as shown in Fig.~\ref{fig:vision}.

Therefore, atom-photon interfaces with a long cavity length $L$, along with high cooperativity and a small footprint, have the potential for channel multiplexing while accommodating hundreds of atoms in each cavity for time-multiplexed entanglement generation.
Satisfying such conditions with free-space cavities can pose challenges owing to diffraction, since high cooperativity and large atom capacity require large mirrors, limiting the number of cavities to be installed in a single vacuum chamber.
Furthermore, the free-space cavity mode features a broad tail, as illustrated in Fig.~\ref{fig:nanofiber}(a), particularly away from the focal plane.
As atoms placed in these regions will experience finite coupling to the cavity, the broad tail region cannot be populated with an atom array, thus potentially consuming a large spatial region and limiting the module-level qubit number.

In contrast to free-space cavities, the waveguide mode of nanophotonic devices~\cite{Goban2014, Thompson2013, Hood2016, Gonzalez2024} exhibits fundamentally different properties because of the inherently small mode area. 
Thus, these systems are compatible with the integrated implementation featuring multiple channels in a single module, as illustrated in Fig.~\ref{fig:nanofiber}(a). 
Narrow-band waveguide mirrors, such as fiber-Bragg gratings (FBGs)~\cite{Kato2015, Ruddell2020}, further, allow wavelength-multiplexed cavities for scalable networking without significant spatial requirements~\cite{supp}.

However, despite significant recent progress in their design and fabrication, nanophotonic waveguides suffer from finite propagation loss, which so far has limited demonstrations to relatively short micrometer-length cavities~\cite{Tamara2021,Polnop2020}.
Such short cavities can be suitable for high-rate operations with a small number of embedded emitters~\cite{Evans2018} and stationary neutral-atom qubits trapped nearby~\cite{Thompson2013,Tiecke2014,Tamara2021}; however, they are insufficient to realize the required large atom capacity ($N \gtrsim 200$, requiring $\sim$~1 mm) for time multiplexing, which is crucial to achieve high HEG throughput in the presence of slow atom transport and state preparation~\cite{supp, Huie2021, Li2024}.
To address this issue, we propose an alternative approach using nanofiber optical cavities~\cite{Kato2015, Kato2019, Nayak2019, horikawa2024}.

\subsection{Nanofiber optical cavities}\label{sec:nanofiber}

Nanofiber optical cavities~\cite{Kato2015, Nayak2019, horikawa2024} feature negligible propagation loss in the waveguide region to simultaneously achieve millimeter-long cavities and high cooperativity~\cite{Ruddell2020, horikawa2024, supp}, thus satisfying key requirements for a scalable multiplexed atom-photon interface.
As illustrated in Fig.~\ref{fig:nanofiber}(b), the nanofiber cavity consists of
(i) a nanofiber region with a sub-wavelength diameter, where atoms are evanescently coupled to a guided mode, and
(ii) pair(s) of FBG mirrors outside the tapered region forming Fabry-P\'erot cavities.
The FBG mirror consists of periodic variations in the refractive index along the optical fiber, which reflects target wavelengths with a peak reflectivity exceeding 99.9\% within the $\sim 100$-GHz wide stopband while transmitting other wavelengths with losses below 0.02\%~\cite{Kato2022}.
Additionally, nanofiber optical cavities are also thermally tunable for cavity locking and in situ external coupling optimization.

To interface tweezer-trapped atoms to the cavity mode, atoms are loaded into an optical tweezer array in free space and subsequently transported to standing-wave traps, which are created by the interference pattern of an optical tweezer and its reflection from the nanofiber surface~\cite{Thompson2013, Tamara2021, Menon2023}.
Standing-wave trapping ensures stable atom-photon coupling because the interference pattern is referenced to the fiber surface~\cite{Nayak2019}.
Snapshots of the trap potentials during transport between free-space and standing-wave traps are shown in Fig.~\ref{fig:nanofiber}(b). 
With adiabatic transport along the $x$ direction, potentially on the order of hundreds of microseconds, a single atom at the trap minimum (green circle) can be transported to the nearest site of the standing-wave trap \cite{Thompson2013,Tamara2021}.
While the transfer to the free-space tweezer is slower than fully free-space atom transport, the time cost is sufficiently absorbed by the time-multiplexed operations with $\sim 200$ atoms to achieve a similar HEG rate to free-space cavities (Fig.~\ref{fig:time-mux}).

Among the laser-coolable atom species, ytterbium atom qubits are particularly suited to nanofiber devices because of their field insensitivity and direct access to telecom-band transitions from metastable states \cite{Covey2019}, along with the possibility of exquisite nuclear-spin qubit control~\cite{Ma2022}.
Nanofiber devices compatible with other atom species such as strontium~\cite{Kestler2023}, rubidium~\cite{Lee2015, Gupta2022} and cesium~\cite{Vetsch2010, Lacroute2012, Goban2012, Beguin2014, Kato2015, Nayak2019}, can be readily manufactured by optimizing the diameter for the respective optical transitions~\cite{Li2024perspective}.
For state-of-the-art intrinsic finesse $\mathcal{F} = 4600$ of nanofiber FBG cavities~\cite{horikawa2024}, strong atom-photon coupling with $C_{\mathrm{in}} > 100$ is expected for the ${}^{3}P_{0} \leftrightarrow {}^3 D_{1}$ transition of ${}^{171}$Yb atoms at 1389 nm (Fig.~\ref{fig:nanofiber}(b)).
See also Supplementary Material for details of possible implementations of the protocols~\cite{supp}.

Further notable features of the nanofiber cavity include its seamless fiber coupling for photon routing, its scalability to long and homogeneous waveguides, and its flexibility towards wavelength multiplexed operations, all while maintaining strong atom-photon coupling.
A long cavity is possible because a majority of photon losses in these cavities come from finite FBG and taper loss currently at the 0.03\% level, with a negligible contribution from the nanofiber region; i.e., there is negligible waveguide propagation loss~\cite{Ruddell2020, Kato2022}.
Indeed, a millimeter-long nanofiber cavity with an intrinsic finesse of $\mathcal{F} = 4600$ has already been demonstrated, which is sufficient to accommodate more than two hundred atoms~\cite{horikawa2024}.

\begin{figure}[t]
    \centering
    \includegraphics[width=3.4in]{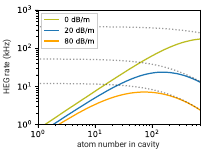}
    \caption{Effect of propagation loss on the time-multiplexed HEG\@. 
    HEG rates are plotted for varying cavity length $L$, containing $(L-\mathrm{1 mm})/dL$ atoms, where $dL=3\times\SI{1.39}{\micro \meter}/n_\mathrm{eff}$, where the effective refractive index $n_\mathrm{eff}=1.07$, is the realistic separation between tweezer sites and the 1~mm spatial overhead comes from the effective length of FBG mirrors $\sim 0.5$~mm~\cite{Barmenkov2006}.
    Cavity loss $\kappa_{\mathrm{in}}$ includes both FBG loss (0.03\%) and waveguide loss (shown as legends). Cavity parameters $\tau$ and $\kappa_{\mathrm{ex}}$ are optimized for respective loss values.
    Dotted lines are the maximum rate $p_e^2/4\Delta t$ for each waveguide loss value with conservative pulse-pulse separation $\Delta t = 10 \tau$. See also Supplementary Material for details~\cite{supp}.
    }
    \label{fig:prop_loss}
\end{figure}

To further illustrate the importance of a small propagation loss for time-multiplexed HEG, we show in Fig.~\ref{fig:prop_loss} the estimations of the time-multiplexed HEG success rate for various loss values, including propagation and mirror loss.
Here, the cavity parameters ($\tau$, $\kappa_\mathrm{ex}$) are optimized at each cavity length $L$ for the length-dependent parameter set $(g(L),\ \kappa_\mathrm{in}(L),\ \gamma)$, to maximize the HEG success rate with Gaussian-shaped photon wavepacket (see Supplemental Material for details~\cite{supp}).
In contrast to the low-loss waveguide with $g(L) > \kappa_\mathrm{in}(L)$ operating at $\tau \sim 1/g$ for the optimal HEG rate, lossy waveguides require significantly slower operations for large $L$ to maintain sufficient $p_e$~\cite{supp}, resulting in a lower HEG rate.

Wavelength multiplexing is possible by incorporating multiple FBG pairs into the fiber region. 
For ${}^{171}$Yb atoms, multiple telecom-band transitions are available for photon generation, potentially multiplexing to at least a few colors for the simultaneous channel use, for example, ${}^{3}P_{0} \leftrightarrow {}^3 D_{1}$, ${}^{3}P_{1} \leftrightarrow {}^3 D_{2}$, and ${}^{3}P_{1} \leftrightarrow {}^3 D_{1}$ transitions, at 1389, 1480 and 1539~nm, to improve the HEG rate beyond the limit of a single channel.

\section{Conclusion and outlook}
\label{sec:conclusion}

In this paper, we discussed the neutral-atom FTQC architecture and identified the need for a modular approach to tackle scalability, along with the stringent throughput requirement it sets on the atom-photon interfaces, which is beyond the state-of-the-art.
As a potential solution to deliver a high inter-module entanglement generation rate, we propose a nanofiber cavity for time- and channel-multiplexed HEG operations.
Although the discussion is based on the near-term specifications of neutral-atom logical processors, the network requirements are expected to remain similar or become more stringent as the system improves and scales up.

Below, we conclude with a few perspectives for tackling the networking bottleneck of neutral-atom QPU and the fault-tolerant protocols necessary to fully utilize the multiprocessor quantum processing cluster enabled by high-bandwidth photonic links.
In Sec.~\ref{sec:outlook_entanglement}, we discuss future directions and potential challenges in entanglement generation using nanofiber optical cavities.
In Sec.~\ref{sec:outlook_ftqc}, we also clarify critical points to be addressed in further optimization of fault-tolerant protocols and computational architectures to realize the multiprocessor fault-tolerant quantum computer based on our approach.
Finally, in Sec.~\ref{sec:repeater}, we provide an outlook on the applications of the techniques introduced here to long-distance quantum communication, which opens the possibility of various types of quantum information processing beyond FTQC\@.

\subsection{Practical considerations for entanglement generation}
\label{sec:outlook_entanglement}
We illustrated the attainable HEG rate with time-multiplexing in Sec.~\ref{sec:fast_heg} using simplified arguments that capture the essential scalings; however, there exists room for further improvements.
For example, our assumption on photon pulse separation $\Delta t$ was relatively conservative with $\Delta t = 10 \tau \sim \SI{1}{\micro \second}$ so that adjacent pulses have a negligible overlap of $10^{-5}$.
The truncation of the Gaussian wavepacket at such a far tail results in complete isolation of pulses with only $10^{-6}$ reduction in the probability of photon generation~\cite{Utsugi2022}.
Thus, a thorough analysis of the source of infidelity and the optimization of the HEG operation may lead to a decrease in $\Delta t$ for an improved HEG rate limit with a single nanofiber device.
Moreover, while we have analyzed HEG using single photons, it would be important to compare such discrete-variable (DV) protocols with continuous-variable (CV) protocols using coherent light~\cite{PhysRevLett.96.240501,Ladd_2006,PhysRevA.78.062319,PhysRevLett.101.040502,PhysRevA.80.060303,PhysRevA.85.060303,PhysRevA.105.062432}.
The optimal trade-off between the rate of entanglement generation and the fidelity of the entangled states generated depends on the entanglement distillation protocols~\cite{Deutsch1996,Bennett1996,Nickerson2013,Krastanov2019}, which should be investigated in more detail.

Moreover, a number of practical challenges need to be addressed to fully exploit the high-rate HEG with a nanofiber atom-photon interface.
A potential problem to be addressed, for example, is the reduction in the coherence time of neutral atoms near the surface of nanofibers~\cite{Tamara2021, Hummer2019}, which may be mitigated by cooling sequences, echo sequences~\cite{Tamara2021} and zoned operations to keep the atoms in the cavity mode for a shorter duration~\cite{supp}.
The effect of nearby ($\lesssim \SI{100}{\micro \meter}$) dielectric charging on Rydberg gate operations~\cite{Ocola2024} can affect local computing in neutral-atom modules, for which careful spatial arrangement of fiber devices and the entangling zone is expected to suffice~\cite{Ocola2024}; the nanophotonic interfaces should be placed at least 100 $\mu$m away from the Rydberg zone to minimize the effects on Rydberg gates, while qubit storage can be near the nanophotonic devices for efficient use of the space.

\subsection{Optimization of fault-tolerant protocols and architectures for neutral-atom multiprocessor fault-tolerant quantum computers}
\label{sec:outlook_ftqc}
For establishing quantitative technological goals for scalable fault-tolerant quantum computers based on criteria~(i)--(vii) shown in Sec.~\ref{sec:criteria}, determining the requirements for the physical error rates and the number of qubits in each module is necessary.
The actual requirements of the modules depend on the details of computational architectures that bridge the physical operations on the atoms and the fault-tolerant protocols to be implemented.

At the physical level, the best candidates for the codes to be used in each module are those with a high threshold to minimize the demands of physical devices, e.g., a $C4/C6$ architecture with a few concatenation levels~\cite{Yoshida2024, Knill_2005, Criger2016}.
Once the logical error rate is suppressed within each module, we can leverage more recent constant-space-overhead protocols for FTQC to reduce the overhead of physical qubits per logical qubit.
This can be achieved by using codes with multiple logical qubits, such as quantum Hamming codes~\cite{Yamasaki2024, Yoshida2024} and non-vanishing-rate quantum LDPC codes~\cite{Gottesman2014, Fawzi2018}.
In terms of physical implementation, the code-concatenation approach offers advantages in modularity because a finite-size code used at each concatenation level serves as an abstraction layer in its implementation~\cite{Yamasaki2024,Yoshida2024}; by properly abstracting the modules of neutral-atom QPUs based on the codes to be concatenated, it is possible to recursively define the finite set of required operations for each module, simplifying the implementation.
On the other hand, the quantum LDPC code approach requires increasing the code size in a code family to improve the logical error rate, and thus on a large scale, even a single block of quantum LDPC code may need to be divided into multiple modules, which necessitates the appropriate combination of operations within and between modules.
Still, careful design of quantum LDPC codes and fault-tolerant protocols can indeed improve the implementability of the protocols~\cite{PhysRevLett.129.050504,Xu2024,Bravyi2024}.

\subsection{Long-distance quantum communication}
\label{sec:repeater}
The highly multiplexed atom-photon interface architecture discussed in this paper is also advantageous for long-distance quantum communication, including quantum key distribution~\cite{Panayi2014}, quantum repeaters~\cite{Azuma2023}, and entanglement-assisted communication protocols~\cite{Bennett1999}. 
Time multiplexing, in this case, also reduces the time cost of the classical communication required to transmit the herald signal to the QPU after photon detection, which takes, for example, 100~$\mu$s for a 20~km internode distance~\cite{Huie2021}.
Thus, a similar requirement is expected in the cavity design for the optimum use of the channels for long-distance communication.
The nanofiber cavities provide distinct characteristics, including operations at multiple telecom-band wavelengths with ytterbium qubits~\cite{Covey2019} and seamless connection to the fiber network. 

The integration of FTQC-ready quantum computing modules via high-bandwidth networking offers a new avenue for quantum information processing with communication.
With such integration, various quantum communication tasks can be explored, including error-corrected (second-generation) repeaters~\cite{Jiang2009}, fault-tolerant channel coding~\cite{Christandl2024,Belzig2024}, blind quantum computing~\cite{Fitzsimons2017}, as well as potential demonstration of the advantages of quantum computation in communication scenarios~\cite{meier2023energyconsumptionadvantagequantumcomputation,yamasaki2023advantage}. 

\begin{acknowledgements}
We thank V.~Vuleti\'c for discussions and comments, and S.~Kato, H.~Konishi, S.~Kikura, and in particular K.~N.~Komagata, for providing feedback on the manuscript.
\end{acknowledgements}

\clearpage
\appendix

\setcounter{table}{0}
\setcounter{figure}{0}
\renewcommand{\thetable}{A\arabic{table}}
\renewcommand{\thefigure}{A\arabic{figure}}
\renewcommand{\theHtable}{Supplement.\thetable}
\renewcommand{\theHfigure}{Supplement.\thefigure}

\section*{Supplementary Material}

\section{Multiplexed Bell pair generation rate}
\label{app:heg}

In this section, we discuss the choice of cavity parameters to achieve a high HEG success rate, with a simple Gaussian wavepacket model applicable to both adiabatic and non-adiabatic conditions \cite{Utsugi2022}. 
In Sec.~\ref{sec:cavity_opt}, we observe how the optimum HEG success rate, upper bounded by $p_e^2/2\tau$, scales with varying cavity qualities such as coupling strengths and internal loss, where $p_e$ is the photon generation success probability and $\tau$ is the characteristic pulse width.
In Sec.~\ref{sec:heg_scaling}, we provide an overview of time-multiplexed HEG for multiple emitters in a cavity~\cite{Huie2021,Li2024}.
In Sec.~\ref{sec:length_scaling}, we observe how the HEG success rate scales with the cavity length, which is proportional to the atom capacity of one-dimensional waveguides.
Finally, in Sec.~\ref{sec:zonemux}, we introduce a zoned operation for time-multiplexed HEG that mitigates the time cost of slow atom transport for a robust time-multiplexed HEG operation with nanophotonic cavities.

\subsection{Optimizing cavity parameters for HEG success rate}\label{sec:cavity_opt}
\begin{figure}[t!]  
    \includegraphics[width=3.4in]{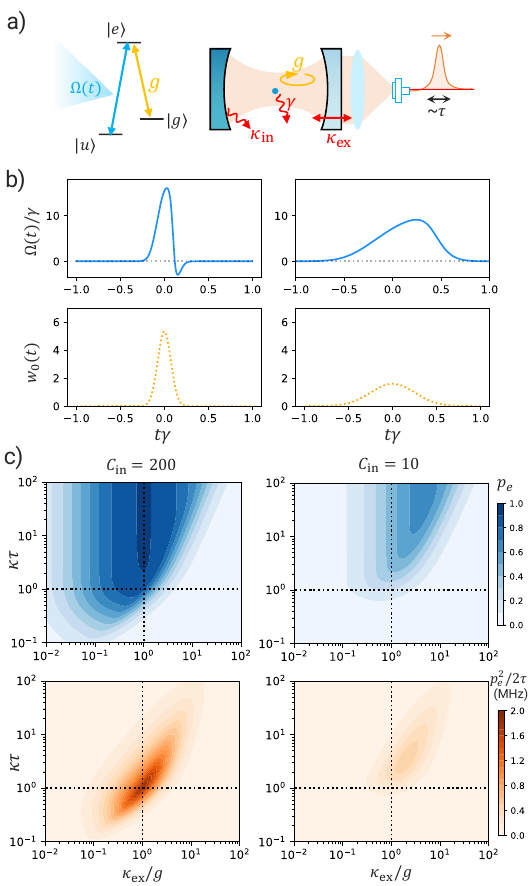}
	\caption{Optimization of the HEG rate within the Gaussian wavepacket model.
        (a) In this model, $|g\rangle \leftrightarrow |e\rangle$ transition of a three-level atom is coupled to the cavity, and $|u\rangle \leftrightarrow |e\rangle$ is driven by a control laser with Rabi frequency $\Omega(t)$.
        The qubit state should span between $|u\rangle$ and another state $|u'\rangle$  (not shown), as well as associated $|g'\rangle$ state where necessary, such that the qubit populations do not mix in the photon generation protocols.
        (b) Examples of waveform $\Omega(t)$ and emitted photon wavepacket are shown for $C=10$ with $\tau = 1.5\tau_c$ (left) and $\tau = 5\tau_c$ (right). 
        Pulse generation with small $\tau$ requires a flip of phase at the end of the driving (the region with $\Omega < 0$) so as to drive the population back into $|u\rangle$ to avoid a long tail in the $w_0(t)$~\cite{Utsugi2022}.
        (c) Photon emission probability $p_e$ (top) and an upper bound $p_e^2/2\tau$ HEG success probability (bottom) are shown for the range of $\tau$ and $\kappa_{\mathrm{ex}}$.
        Horizontal and vertical dashed lines are $\tau=\kappa$ and $\kappa_{\mathrm{ex}} = g$.
    }
    \label{fig:psmax}
\end{figure}

With three-level atoms coupled to a cavity, the generation of photons with a desired temporal wavepacket is possible by a control field with time-dependent Rabi coupling $\Omega(t)$~\cite{Vasilev2010}, as shown in Fig.~\ref{fig:psmax}(a).
For example, the Gaussian pulse generation illustrated in Fig.~\ref{fig:psmax}(b)~\cite{Utsugi2022} is well-suited for our purpose owing to its robustness against temporal mismatches~\cite{Rohde2005}.
The relation between the Gaussian pulse width $\tau$ and the total photon emission probability $p_e$ for a given cavity parameter ($g,\ \kappa_{\mathrm{in}},\ \kappa_{\mathrm{ex}},\ \gamma$) is given in both the adiabatic ($\tau > \tau_c$) and non-adiabatic regimes ($\tau \lesssim \tau_c$), as shown in the top panels of Fig.~\ref{fig:psmax}(c) (blue), where $\tau_c = \max(1/\kappa, \kappa/g^2)$ and $\kappa=\kappa_\mathrm{in}+\kappa_\mathrm{ex}$~\cite{Utsugi2022}. 
Herein, we vary $\kappa_\mathrm{ex}$ and $\tau$ for exemplary sets of cavity parameters ($g,\ \gamma, \kappa_\mathrm{in}$)$/2\pi$  =  (5, 0.25, 0.25) MHz and (5, 0.25, 5) MHz, corresponding to $C_\mathrm{in} = $200, 10.
From the figures, it is clear that a high $p_e$ requires long pulses $\tau>\tau_c$ and an optimal choice of $\kappa_\mathrm{ex}$.

In contrast, optimizing the HEG success rate $p_e^2/2\tau$ leads to a global optimum near $\tau \sim 1/\kappa$ and $\kappa_{\mathrm{ex}} \sim g$ for cavities with high internal cooperativity, as shown by the bottom left panel of Fig.~\ref{fig:psmax}(c).
This is because, for an optimal $\kappa_{\mathrm{ex}}$, a relatively high $p_e$ can be maintained even with short pulses at $\tau \sim 1/\kappa$ (see the top left panel of Fig.~\ref{fig:psmax}(c)), providing an optimal balance between fast photon generation and high $p_e$.
Notably, in the case of nanofiber optical cavities, the FBG mirrors allow precise in-situ tunability of $\kappa_{\mathrm{ex}}$ by thermal tuning~\cite{Kato2019}.
As $\tau$ is by design a controllable parameter through $\Omega(t)$, the optimizations shown in the bottom panes of Fig.~\ref{fig:psmax}(c) can be performed in situ to maximize the performance of the actual parameters of the installed cavity, which may differ from the designed parameters.

\begin{figure}[t!]  
    \includegraphics[width=0.83\linewidth]{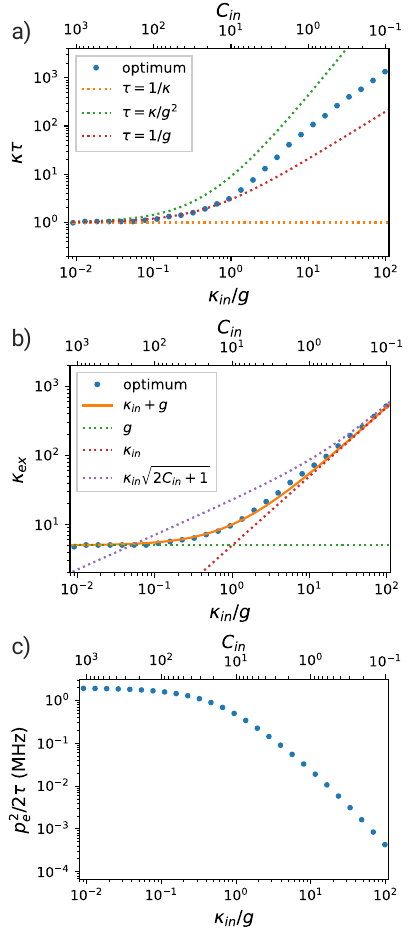}
	\caption{Optimization of cavity parameters for HEG success rates.
            (a) Optimized pulse length $\tau$ is plotted in units of $1/\kappa$ as points, compared to several scaling functions. The points follow $\tau=1/g$ for $g>\kappa_{\mathrm{in}}$.
            (b) Optimized $\kappa_{\mathrm{ex}}$ for HEG operations is plotted, showing scaling consistent with $\kappa_{\mathrm{ex}} \sim g + \kappa_{\mathrm{in}}$. The observed scaling is different from $\kappa_{\mathrm{ex}} = \kappa_{\mathrm{in}}\sqrt{2C_{\mathrm{in}} + 1}$, an optimal value for photon generation probability for $\tau \gg \tau_c$.
            (c) Optimized HEG rate $p_e^2/2\tau$ is plotted, where the factor of two is due to the upper bound of the success probability of Bell basis measurement by photon detection.
    }
    \label{fig:cavity_param_scaling}
\end{figure}

\begin{figure}[t!]
    \includegraphics[width=3.4in]{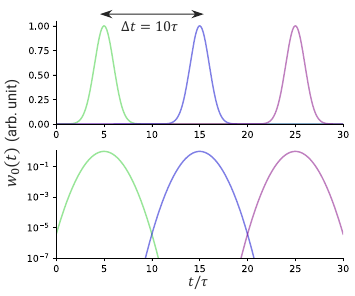}
	\caption{Temporal separation of Gaussian pulse train in linear scale (top) and logarithmic scale (bottom), where different colors indicate pulses from different atoms. With $\xi=10$ ($\Delta t = 10 \tau$), overlap is limited to $< 10^{-5}$.
    }
    \label{fig:pulse_separation}
\end{figure}

The optimal parameters deviate from the aforementioned relationships for smaller $C_{\mathrm{in}}$ (bottom right panel of Fig.~\ref{fig:psmax}(c)) and require longer pulses compared to $1/\kappa$.
For a quantitative analysis of the shift for varying cavity qualities, as shown in Fig.~\ref{fig:cavity_param_scaling}, we plot the scaling of the optimal cavity parameters for HEG operations for varying $\kappa_{\mathrm{in}}/g$, with ($g,\ \gamma$)$/2\pi$  =  (5, 0.25) MHz.
Figures~\ref{fig:cavity_param_scaling}(a) and (b) show the optimal $\tau$ and $\kappa_{\mathrm{ex}}$, respectively, to maximize $p_e^2/2\tau$, which follows scaling $\tau \sim 1/g$ for $\kappa_{\mathrm{in}}/g \lesssim 1$ and $\kappa_{\mathrm{ex}} \sim g + \kappa_{\mathrm{in}}$.
The optimal parameters for the HEG differ from the known optimum for photon generation probability, such as $\kappa_{\mathrm{ex}}=\kappa_{\mathrm{in}}\sqrt{2C_{\mathrm{in}}+1}$~\cite{Goto2019}.
As shown in Fig.~\ref{fig:cavity_param_scaling}(c), for low-loss cavities with $C_{\mathrm{in}} \gtrsim 100$, the upper bound of the HEG rate reaches $\sim 2$ MHz; however, practical considerations for high-fidelity operations are required, such as sufficient temporal separation of pulses to avoid pulse overlaps.
Here, we take a conservative value of $\Delta t = \xi \tau$ with $\xi = 10$, where the pulse-pulse overlap is below $10^{-5}$, as illustrated in Fig.~\ref{fig:pulse_separation}.
The time-bin protocol incurs an additional factor of two in photon generation time overhead and a short additional $\pi$ transition time (absorbed in $t_{\mathrm{init}}$), resulting in a conservative estimate for the HEG rate upper bound $p_e^2/4\Delta t \sim 100$ kHz.
The optimized pulse shape and relaxed conditions for pulse overlap allow further packing of pulse train for smaller $\Delta t$, for the upper bound of a few hundred kHz or $\sim$ MHz HEG success rate to become realistic.
For example, generation of a truncated Gaussian pulse with truncation at $5\tau$ from the peak is possible with only a $10^{-6}$ reduction in photon emission probability $p_e$, and the cost is only $10^{-3}$ for truncation at $2.5\tau$, as shown in Ref.~\cite{Utsugi2022} using numerical solutions of master equation.

\begin{figure*}[t]
    \includegraphics[width=7.0in]{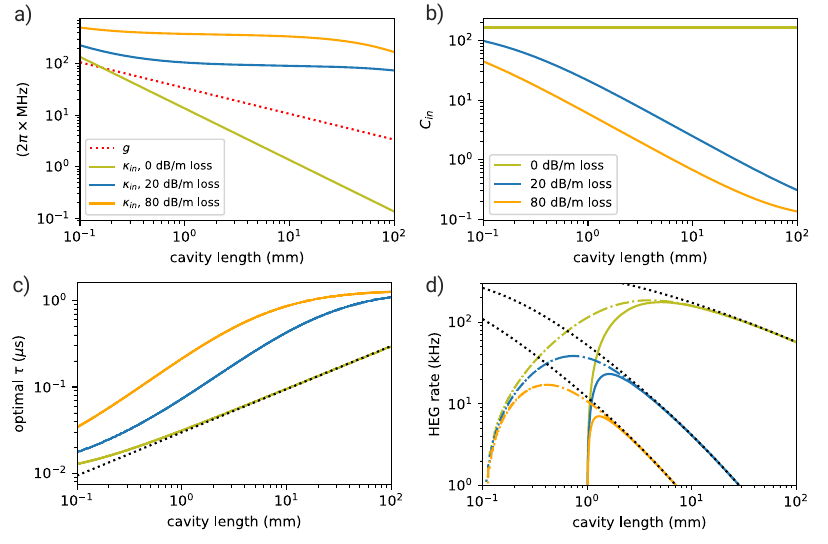}
	\caption{Length dependence of cavity parameters and multiplexed HEG rate.
            (a) Scaling of cavity parameters is shown as a function of cavity length $L$.
            (b) Internal cooperativity is calculated for the parameters in (a).
            (c) Optimum $\tau$ is shown for respective waveguide loss values, compared with $1/g$ (black dotted line).
            (d) Time-multiplexed HEG rate is shown for cavities with respective loss constants. 
            Black dotted lines are the optimal values $p_e^2/4\Delta t$ for each loss constant.
            Solid lines are the rates calculated with effective mirror length of $L_m = 1~\mathrm{mm}$ and dash-dotted lines are $L_m = 100~\mu$m, such that the number of atoms is given by $(L-L_m)/dL$.
    }
    \label{fig:length_scaling}
\end{figure*}

\subsection{Time-multiplexed HEG}\label{sec:heg_scaling}
In Fig.~\ref{fig:time-mux}, we show how the bell pair generation rate scales with the atom number $N$ in a more practical situation, that is, in the presence of a large time cost for atom transport and state initialization $(t_\mathrm{move}, t_\mathrm{init})=$(100~$\mu$s, 20~$\mu$s) and $(t_\mathrm{move}, t_\mathrm{init})=$(1000~$\mu$s, 20~$\mu$s), with the time-multiplexed protocol depicted in Fig.~\ref{fig:time-mux}. 
Essentially, slow atom transport is interleaved by long HEG operations consisting of $M$ repetitions of atom initialization and $N$ sequential photon generation trials~\cite{Huie2021}.
The cumulative number of Bell pairs generated over $M$ repetitions is denoted by $N_M = \sum_{i=1}^M N_i p_e^2/2$, where $N_i=N(1-p_e^2/2)^{i-1}$.
The total time cost is $t_M = t_\mathrm{move}+M t_\mathrm{init}+2\Delta t \sum_{i=1}^M N_i$.
Thus, the averaged Bell pair generation rate is given by $R_M = N_M/t_M$~\cite{Huie2021, Li2024}.
The repetition number $M$ must be optimized to maximize $R_M$, which occurs at $M\sim 5$ for the parameters used in Fig.~\ref{fig:time-mux}(c).
A large $N$ is required to saturate the rate at $\sim p_e^2/4\Delta t$ by satisfying $t_{\mathrm{move}},\ M t_\mathrm{init} \ll 2\Delta t \sum_i^M N_i$. 
In practice, as shown in Fig.~\ref{fig:time-mux}(c), $N\sim200$ is enough to reach the practical HEG rate at 90\% of the upper bound.

\subsection{Nanophotonic cavity: length dependence}\label{sec:length_scaling}
While a large $g$ attained by short cavities is attractive for fast operation, the necessity of time multiplexing calls for competing characteristics of longer cavities to accommodate hundreds of atoms in a cavity mode.
The scaling of cavity parameters for varying cavity lengths $L$ is given by $g(L) \propto 1/\sqrt{L}$ and $\kappa_{\mathrm{in}}(L), \kappa_{\mathrm{ex}}(L) \propto 1/L$ with negligible waveguide (propagation) loss~\cite{Reiserer2015}, which is the case for nanofibers.
In this case, internal cooperativity is independent of cavity length, as dependence cancels out between $\kappa_{\mathrm{in}}$ and $g^2$.
In the presence of waveguide loss, $\kappa_{\mathrm{in}}$ becomes nearly length-independent; see Fig.~\ref{fig:length_scaling}(a) for waveguide loss constants 0~dB/m, 20~dB/m and 80~dB/m\@.
This results in a length dependence of $C_{\mathrm{in}}$, as shown in Fig.~\ref{fig:length_scaling}(b).

For each $(g(L), \kappa_{\mathrm{in}}(L))$, we find optimal $\tau(L)$ and $\kappa_{\mathrm{ex}}(L)$ to maximize the HEG rate and plot optimal $\tau$ and the time-multiplexed entanglement generation rate in Figs.~\ref{fig:length_scaling}(c) and~(d), for different loss constants of the waveguide, with the number of atoms in the cavity $N=(L-L_m)/dL$ and two values of $L_m = 1~\mathrm{mm}$ and $100~\mu$m.
The results shown in Fig.~\ref{fig:length_scaling}(d) with $L_m = 1~\mathrm{mm}$, shown as solid lines, are used for Fig.~\ref{fig:prop_loss} in the main text, with the horizontal axis converted into the atom number.

\begin{figure*}[t!]
    \centering
    \includegraphics[width=7.0in]{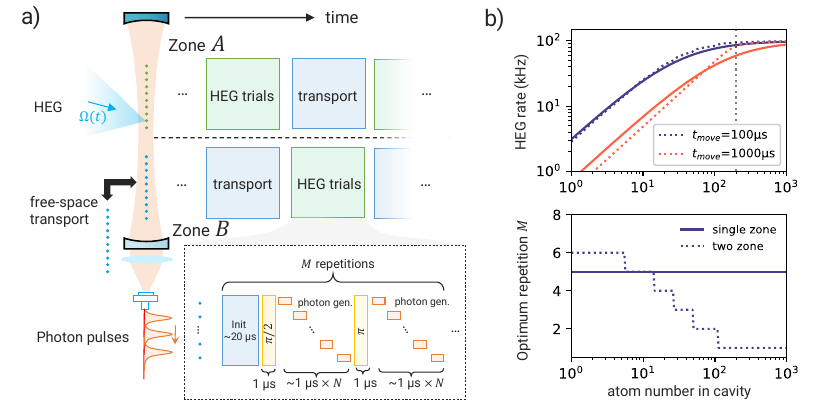}
    \caption{Zoned time-multiplexing for high-rate HEG in the presence of slow atom transport. 
            (a) The concept of zoned time multiplexing for high-rate HEG is illustrated. The cavity mode is partitioned into two regions, Zone A and B, where atoms operate different sequences of atom transport and HEG trials.
            The time cost of slow transport is removed by parallelizing with simultaneous HEG trials, at the cost of half the atom number in each zone for efficient time multiplexing.
            (b) Simulated performances of single (solid) and dual (dotted) zone operations are plotted. 
            Approximately 50\% improvement in the rate is observed for slow transport and relatively large numbers of atoms.
            The atom number required for saturating at 90\% of the best rate (100 kHz), in the slow transport case, changes from $\sim 1000$ to $\sim 200$, which is an improvement by a factor of $5$.
            Another characteristic is the smaller number of repetitions $M$ to achieve the best rate, which reduces the detrimental effect of storing already entangled atoms in the cavity mode while other atoms are emitting photons.
    }
    \label{fig:zone_mux}
\end{figure*}

\subsection{Zoned operation for time-multiplexed HEG}\label{sec:zonemux}
By partitioning the cavity mode into two regions $A$ and $B$, and performing atom transport and entanglement generation operations simultaneously between the two zones, the time-multiplexed HEG rate can be improved for the case of slow atom transport, as shown in Fig.~\ref{fig:zone_mux}(a).
Essentially, the time cost changes from $t_M = t_\mathrm{move}+M\cdot t_\mathrm{init}+\Delta t \sum_{i=1}^M N_i$ to $t_M = \max\left\{ t_\mathrm{move},\ M t_\mathrm{init} + \Delta t \sum_{i=1}^M N_i\right\}$ while the atom number $N$ in each zone is half of the total number $N$.
This is most effective when $t_\mathrm{move} \sim M t_\mathrm{init} + \Delta t \sum_{i=1}^M N_i$, which is the case for, e.g., $t_{\mathrm{move}} = 1000$~$ \mu$s and $\Delta t \sim 1$~$\mu$s, as shown in Figs.~\ref{fig:time-mux} and~\ref{fig:zone_mux} (b).
A notable characteristic of zoned operation is that the optimum $M$ is small, reaching $M=1$ for sufficiently large $N$ (see Fig.~\ref{fig:zone_mux}(b), bottom panel). 
This characteristic is beneficial in ensuring high fidelity while maintaining a high HEG rate because keeping already entangled atoms in the cavity for multiple $M$ over hundreds of $\mu$s results in rapid accumulation of errors.

\section{${}^{171}$Yb atom implementations}
\begin{figure}[t!]
    \centering
    \includegraphics[width=3.5in]{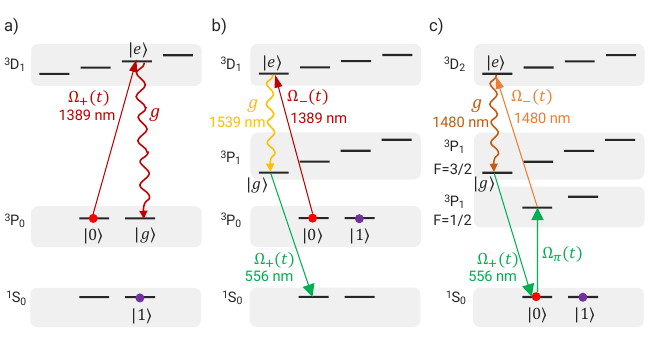}
    \caption{Telecom-band single-photon generation level schemes for ${}^{171}$Yb atoms coupled nanofiber cavities.
    (a) A 1389 nm photon is generated, for example, by starting at ${}^{3}\mathrm{P}_{0}$ metastable state, driving a transition between $|{}^{3}\mathrm{P}_{0}, m_F=-1/2\rangle$ and $|{}^{3}\mathrm{D}_{1},m_{F'}=+1/2\rangle$ with a time-dependent pulse and collecting cavity-enhanced emission of linearly-polarized photon. 
    (b) A 1539 nm photon generation is possible by driving $\sigma_-$ transition to ${}^{3}\mathrm{D}_{1}$ manifold with cavity tuned for ${}^{3}\mathrm{P}_{1} \leftrightarrow {}^{3}\mathrm{D}_{1}$ $\pi$ transition, while applying $\Omega_+$ driving between ${}^{3}\mathrm{S}_{0} $ and $ {}^{3}\mathrm{P}_{1}$ to coherently transfer the population to ground state, in a similar manner to the scheme described in Ref.~\cite{Huie2021}.
    (c) A 1480 nm photon generation protocol closely follows Ref.~\cite{Huie2021}, with opposite order of population transfers and cavity tuned to the transition between $|{}^{3}\mathrm{P}_{1}, m_F=-3/2\rangle$ and $|{}^{3}\mathrm{D}_{2}, m_{F'}=-3/2\rangle$. 
    }
    \label{fig:yb_atom}
\end{figure}

As concrete examples for implementing the above protocols, we outline a few possible level schemes for telecom-band photon generations with ${}^{171}$Yb atoms coupled to the nanofiber cavity.
Previously proposed protocols include circularly polarized photon generation with 1539 nm~\cite{Li2024} and 1480 nm transitions~\cite{Huie2021}.
Here, we consider a generation of a linearly polarized photon, which is more efficiently coupled to the waveguide cavity mode.
The 1389 nm photon generation is possible by an effective three-level system of $|{}^{3}\mathrm{P}_{0}, m_F=\pm1/2\rangle$ and $|{}^{3}\mathrm{D}_{1}, m_F=+1/2\rangle$, where other states in the ${}^{3}\mathrm{D}_{1}$ manifold can be detuned by applying a static magnetic field, as illustrated in Fig.~\ref{fig:yb_atom} (a).
The initial qubit population can be prepared in a coherent superposition between $|{}^{3}\mathrm{P}_{0}, m_F=-1/2\rangle$ (red circle in Fig.~\ref{fig:yb_atom}) and $|{}^{1}\mathrm{S}_{0},m_F=+1/2\rangle$ (purple circle in Fig.~\ref{fig:yb_atom}), where driving the $\sigma_\pm$ clock transitions $|{}^{1}\mathrm{S}_{0},m_F=\mp1/2\rangle \leftrightarrow |{}^{3}\mathrm{P}_{0},m_F=\pm 1/2\rangle$ before the second photon generation allows generation of atom-photon entanglement with time-bin encoding of the photon at a 1389 nm wavelength.
Alternatively, a 1539 nm photon can be generated by coherent transfer of the population as shown in Fig.~\ref{fig:yb_atom} (b). 
Here, in addition to the three-level structure, an additional state in the ground-state manifold is involved to ensure that the atoms are decoupled from the cavity after photon generation, similar to the scheme in Refs.~\cite{Menon2020,Huie2021}.
For this protocol, the spin flip between photon generations involves $\pi$ pulses for both the ${}^{3}\mathrm{P}_{0}$ metastable manifold and the ${}^{1}\mathrm{S}_{0}$ ground state, to avoid spin-flip (Pauli $X$) errors from the mixing of populations during photon generation.
Finally, 1480 nm photon generation closely follows Ref.~\cite{Huie2021} in an opposite direction as illustrated in Fig.~\ref{fig:yb_atom} (c).

%\bibliography{refs}

%

\end{document}